\documentclass[useAMS,usenatbib]{mn2e}
\usepackage{graphicx}
\title[HR\--IR spectra of NGC6624 \& NGC6569]{High resolution near IR spectra of NGC~6624 and NGC~6569\thanks{Data presented 
herein were obtained at the W.~M.~Keck 
Observatory, which is operated as a scientific partnership among the California Institute 
of Technology, the University of California, and the National Aeronautics and Space Administration. 
The Observatory was made possible by the generous support of the W.~M.~Keck Foundation. 
}}
\author[E. Valenti, L. Origlia \& R.~M. Rich]{E. Valenti$^{1}$\thanks{E-mail:
evalenti@eso.org (EV)}, L. Origlia$^{2}$ \& R.~M. Rich$^{3}$\\
$^{1}$European Southern Observatory, Karl Schwarzschild\--Stra\ss e 2, D\--85748 Garching bei M\"{u}nchen, Germany.\\
$^{2}$INAF - Osservatorio Astronomico di Bologna, via Ranzani 1, I-40127 Bologna, Italy\\
$^{3}$Department of Physics and Astronomy, University of California at Los Angeles, Los Angeles, CA 90095 - 1562, US}
\begin{document}

\date{}

\pagerange{\pageref{firstpage}--\pageref{lastpage}} \pubyear{2010}

\maketitle

\label{firstpage}

\begin{abstract}
We present the first abundances analysis based on high\--resolution infrared (IR) echelle spectra of NGC~6569 and NGC~6624, 
two moderately reddened globular clusters located in the outer bulge of the Galaxy. 

\noindent
We find [Fe/H]=\--0.79$\pm$0.02~dex and [Fe/H]=\--0.69$\pm$0.02~dex for NGC~6569 and NGC~6624, respectively and an average 
$\alpha$\--elements enhancement of
$\approx$+0.43$\pm$0.02~dex and +0.39$\pm$0.02~dex, consistent with previous measurements on other metal\--rich Bulge clusters.
We measure accurate radial
velocities of $\rm <v_r>=\--47\pm 4~km~s^{-1}$ and $\rm <v_r>=+51\pm 3~km~s^{-1}$ and velocity dispersions of $\rm \approx 8~km~s^{-1}$
and $\rm \approx6~km~s^{-1}$ for NGC~6569 and NGC~6624, respectively.

\noindent
Finally, we find very low $^{12}C/^{13}C$ isotopics ratio ($\leq$7 in NGC~6624 and $\approx$5 in NGC~6569),
confirming the presence extra\--mixing mechanisms during the red giant branch evolution phase.

\end{abstract}


\section{Introduction}
The new generation of high\--resolution IR spectrographs showed its tremendous potential 
to study both distant and obscured stellar populations. 
Particularly in the case of  heavily reddened regions such as the Galactic bulge and center, IR spectroscopy offers the best, 
and sometimes a unique, approach to measuring the composition of the old stellar populations.

In the last few years, we have started a high\--resolution spectroscopic survey of the Galactic bulge in the near\--IR
by using NIRSPEC, a high\--throughput IR echelle spectrograph at the Keck Observatory \citep{mclean98}. 
H\--band ($\rm 1.5 \-- 1.8 \mu$m) spectra of bright giants in the Bulge globular clusters and field population are ideal for 
detailed abundance analysis of Fe, C, O and other $\alpha$\--elements, using the approach of synthesizing the entire spectrum. 
The abundance distributions in the cluster and field populations are crucial in constraining the history of Bulge formation 
and chemical enrichment \citep{mcwil97}.

We have used this method to derive abundances for ten globular clusters in the inner and outer Bulge regions: 
the resulting abundances for NGC6553 and Liller~1 are given in \citet{ori02}, for Terzan~4 and Terzan~5 in \citet{ori04}, 
for NGC~6342 and NGC~6528 in \citet{ori05a}, for NGC~6539 and UKS~1 in \citet{ori05b}, and for NGC~6440 and NGC~6441 in \citet{ori08}.
We also measured detailed abundances of Bulge M giants in the Baade's window \citep{rich05} and in three inner fields at 
$\rm (l,b)=(0^{\circ},-1^{\circ})$ \citep{rich07},  $\rm (l,b)=(0^{\circ},-1.75^{\circ})$, $\rm (l,b)=(1^{\circ},-2.65^{\circ})$ \citep{rich10}.
We found $\alpha$\--enhancement at a level of a factor between 2 and 3 over the whole range of metallicity spanned by the 
Bulge clusters in our survey, from [Fe/H]$\approx-1.6$ (cf. Terzan~4) up to [Fe/H]$\approx-0.2$ (cf. Terzan~5).

Here we present the high resolution IR spectra and abundance analysis of bright giants in NGC~6569 and NGC~6624, 
two globular clusters of the outer Bulge, located at $\rm (l,b)=(0.48,-6.68)$ and $\rm (l,b)=(2.79,-7.91)$, respectively \citep{harris}.

\begin{figure*}
\includegraphics[width=8.5cm, height=8.cm]{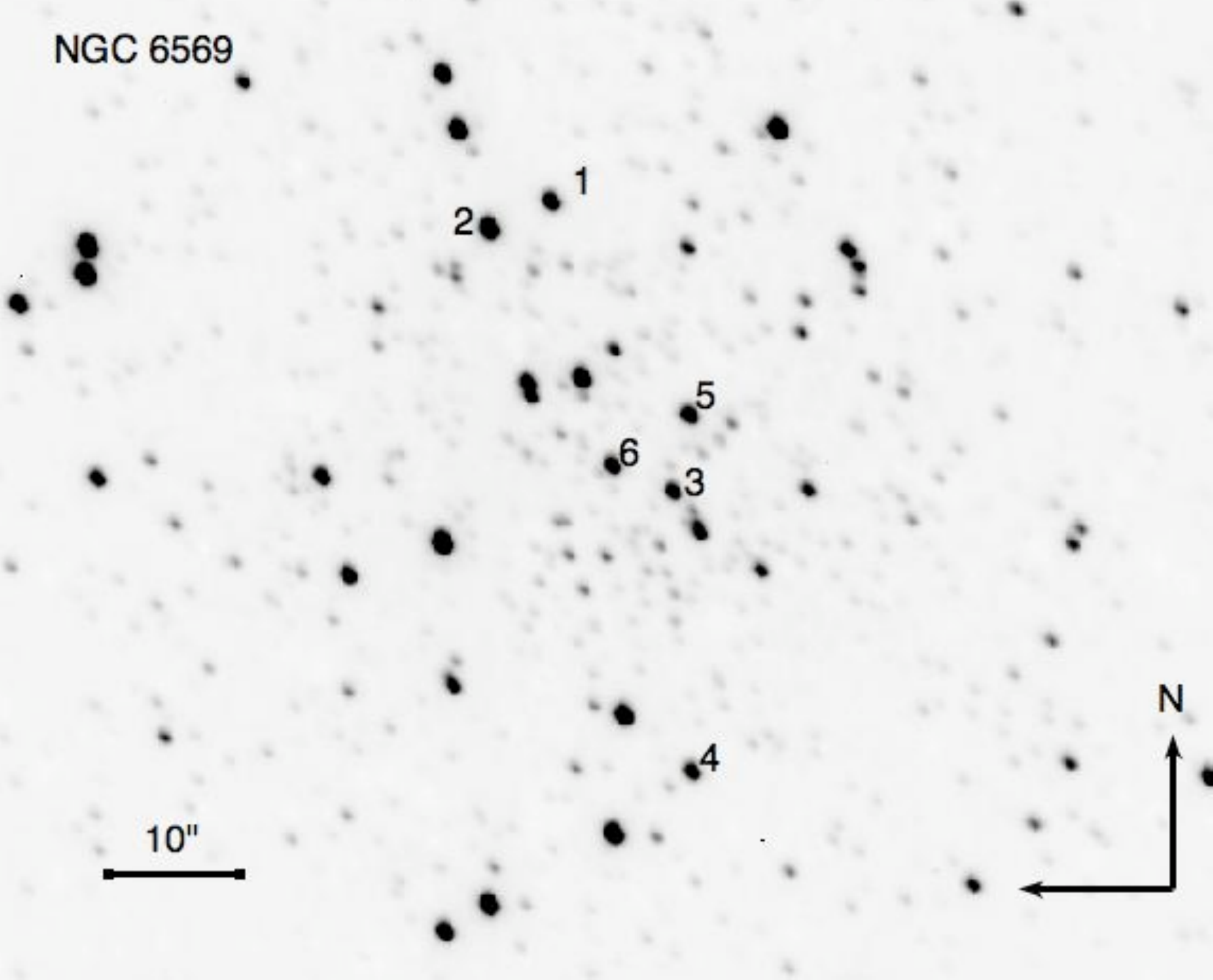}
\includegraphics[width=8.5cm]{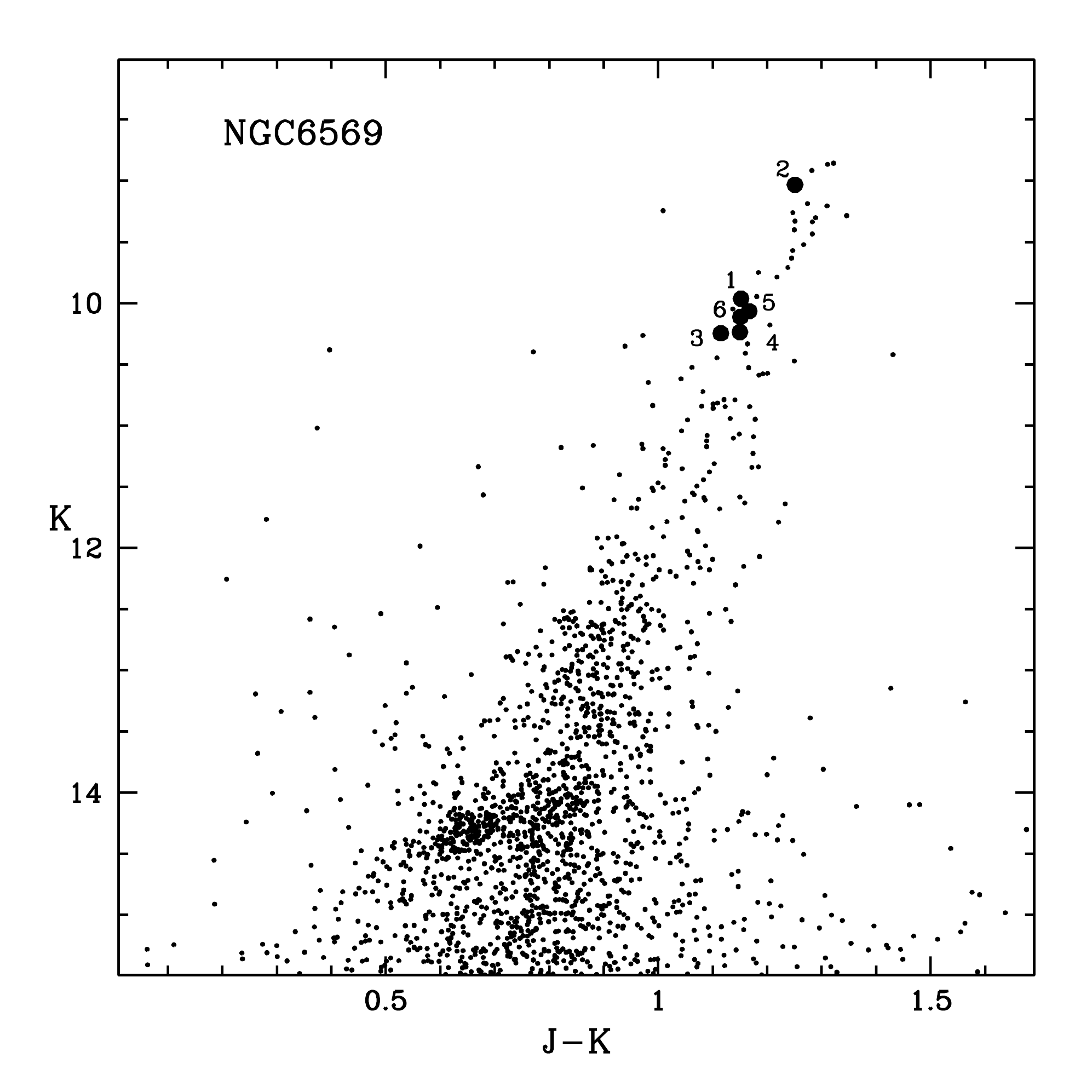}
\caption{H\--band image of the core region (left panel) and the K, J\--K colour\--magnitude diagram 
(right panel) of NGC~6569 \citep{val05}. The stars spectroscopically observed are numbered (cf. Table~1).}
\label{6569data}
\end{figure*}

NGC~6569 is a rather compact cluster located in the Sagittarius region. \citet{zinn85} from integrated photometry obtained 
[Fe/H]$=-0.86$ and E(B\--V)$=0.55$, while from integrated DDO photometry, \citet{bica83} estimated  [Fe/H]$=-0.76$ and E(B\--V)$=0.59$.  
The optical photometric studies by \citet{ort01} and \citet{pio02} suggested a metallicity consistent with that of 47~Tuc. 
The only IR photometric study is that by \citet{val05} who found  [Fe/H]$=-0.85$ and E(B\--V)$=0.49$. 
No high spectral resolution measurements of this cluster exist so far.

In the last few years, NGC~6624 has been subject of several studies aimed at investigating its population of "exotic" objects, 
such as X\--ray binaries \citep[see][ and reference therein]{zdz07} and millisecond pulsars \citep[see][]{scott05}. 
Deep optical HST photometry, down to the main sequence turnoff region, have been presented by \citet{heash00,pio02} finding the cluster to 
be coeval to 47~Tuc. Its similarity to 47~Tuc is also based on the metallicity, in fact \citet{heash00} found [Fe/H]$=-0.63$ 
based on observations of the Ca~II IR triplet lines in giant stars of NGC~6624. More recently, \citet{val04a,val04b} presented a 
detailed analysis of the cluster Red Giant Branch properties based on high resolution IR photometry, finding E(B\--V)=0.28, 
(m-M)$_0$=14.63 and a metallicity like 47~Tuc. However, also in the case of NGC~6624, no high resolution spectroscopic 
measurements have been published so far.

A description of the observations and abundance analysis follows in \S~2, while in \S~3 and \S~4 we
present and discusses our results.  

\begin{table}
\begin{center}
\caption{Coordinates and photometric parameters for the giants observed in NGC~6569 and NGC~6624.}
\label{photparam}
\begin{tabular}{ccccc}
\hline \hline
& & & &  \\
 Star & R.A. & Decl. & (J\--K)$^a_0$ & M$^a_{bol}$\\
& & & & \\
\hline \hline
NGC~6569\--s1 &18:13:39.015 &-31:49:18.88 &0.91 &-3.06\\
NGC~6569\--s2 &18:13:39.392 &-31:49:21.00 &1.01 &-3.85\\
NGC~6569\--s3 &18:13:38.276 &-31:49:41.29 &0.88 &-2.83\\
NGC~6569\--s4 &18:13:38.165 &-31:50:02.95 &0.91 &-2.79\\
NGC~6569\--s5 &18:13:38.179 &-31:49:35.40 &0.93 &-2.94\\
NGC~6569\--s6 &18:13:38.647 &-31:49:39.28 &0.91 &-2.91\\
& & & & \\
NGC~6624\--s1 &18:23:40.129 &-30:21:53.72 &1.04 &-3.76\\
NGC~6624\--s2 &18:23:40.100 &-30:21:47.57 &1.04 &-3.41\\
NGC~6624\--s3 &18:23:41.646 &-30:21:41.19 &1.01 &-3.05\\
NGC~6624\--s4 &18:23:40.635 &-30:21:44.83 &1.01 &-3.12\\
NGC~6624\--s5 &18:23:42.136 &-30:21:26.29 &1.08 &-3.85\\
\hline
\multicolumn{5}{l} {(a) NGC~6569: J, K colours, E(B\--V)=0.49, and (m\--M)$_0$=15.40}\\
\multicolumn{5}{l} {~~~ from \citet{val05}}\\
\multicolumn{5}{l} {~~~ NGC~6624: J, K colours, E(B\--V)=0.28, and (m\--M)$_0$=14.63}\\
\multicolumn{5}{l} {~~~ from \citet{val04a}}\\
\end{tabular}
\end{center}
\end{table}

\section{Observations and abundance analysis}
\begin{figure*}
\includegraphics[width=8.2cm, height=8cm]{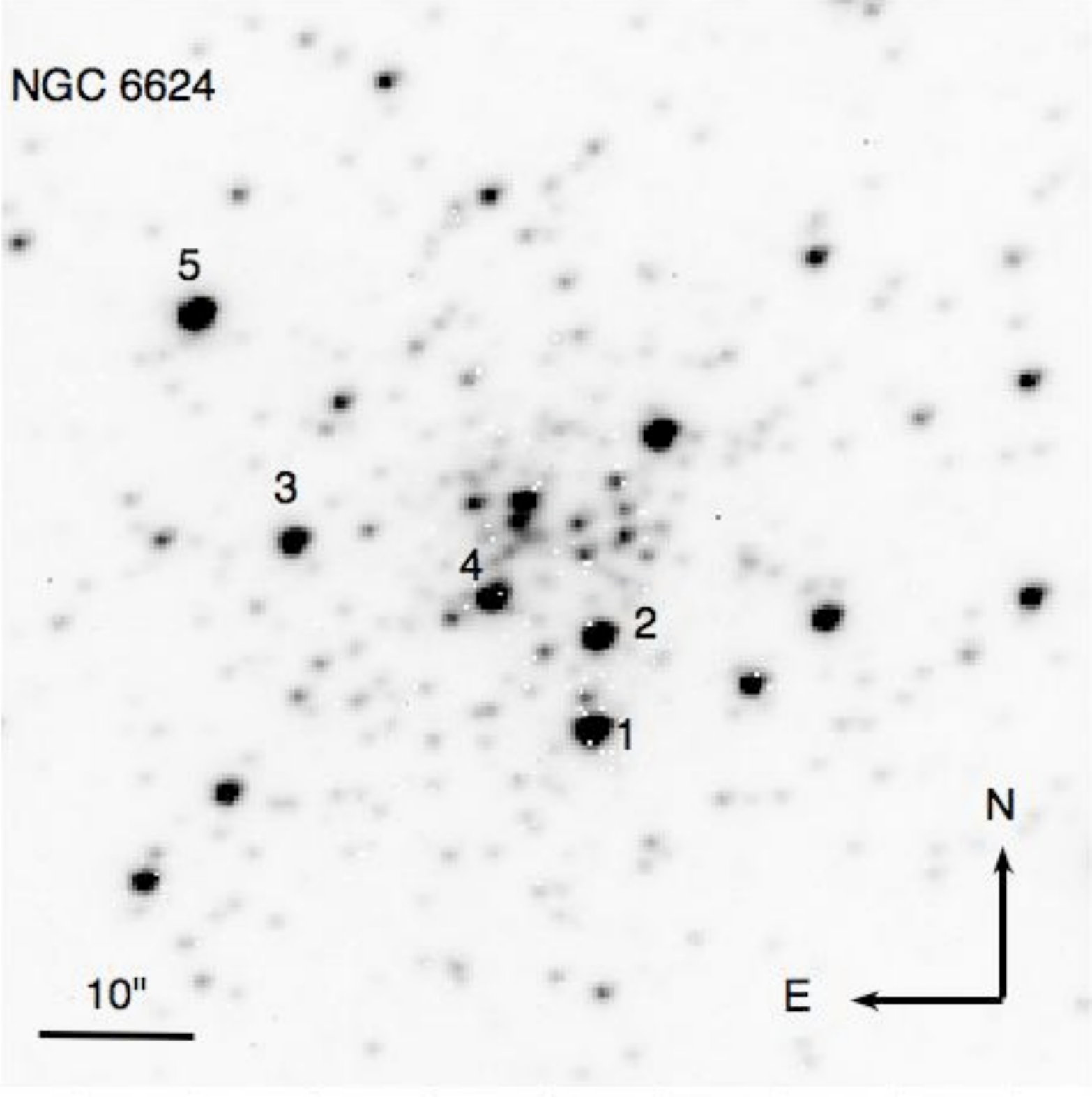}
\includegraphics[width=8.5cm]{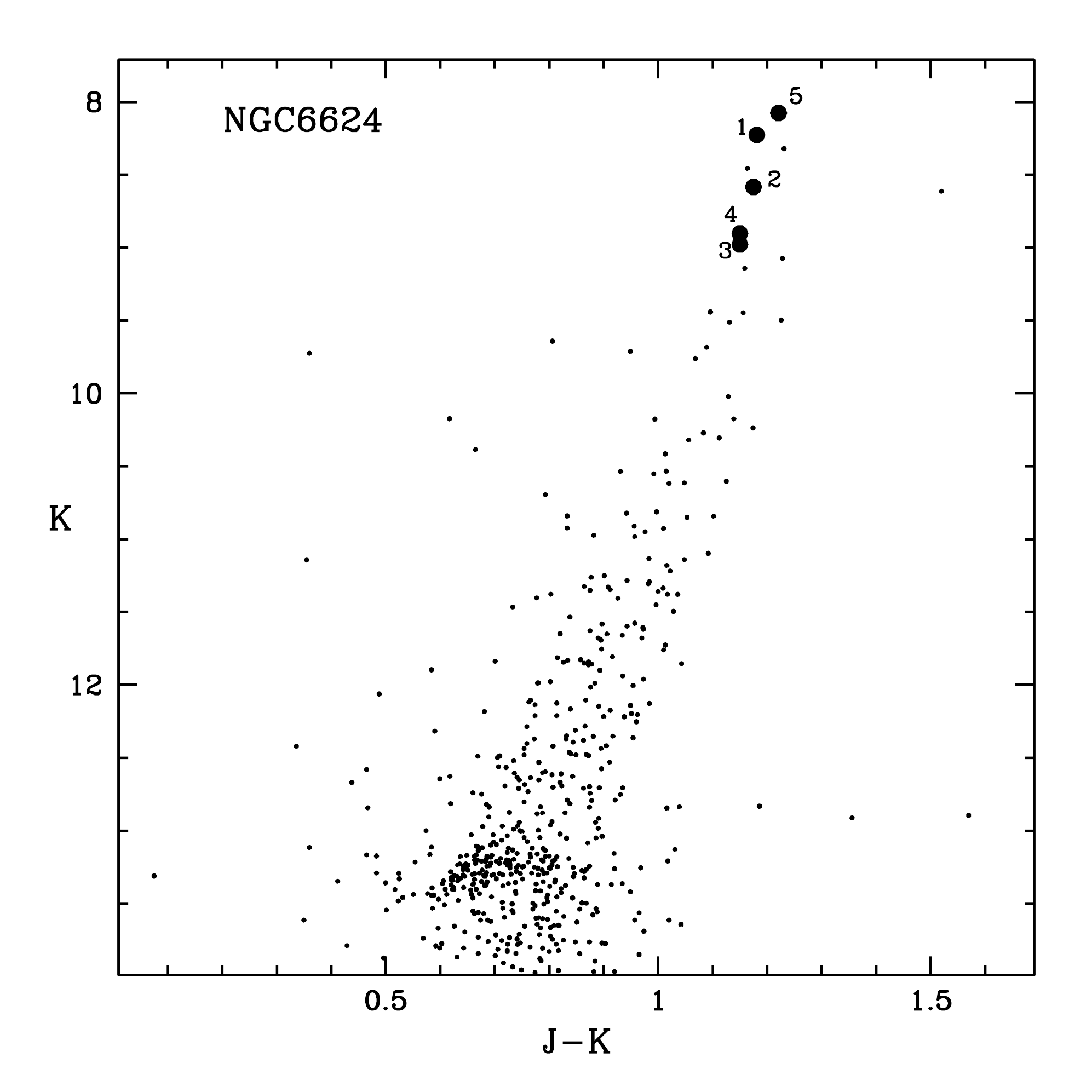}
\caption{H\--band image of the core region (left panel) and the K, J\--K CMD (right panel) of NGC~6624 
\citep{val04a}. The stars spectroscopically observed are numbered (cf. Table~1).}
\label{6624data}
\end{figure*}

Near\--IR, high resolution echelle spectra of bright giants in the core of the Bulge globular clusters NGC~6569 and NGC~6624 have 
been acquired with the IR spectrograph NIRSPEC \citep{mclean98} mounted at the Nasmyth focus of the Keck~II telescope. 
NGC~6569 was observed on May 2006, while NGC~6624 on April 2004, May 2005 and May 2006. 
The high resolution echelle mode, with a slit width of $\rm 0.43\arcsec$ (3 pixels) and a length of $\rm 12\--24\arcsec$ and the 
standard NIRSPEC\--5 setting, which covers most of the $\rm 1.5 \--1.8 \mu$m H\--band, have been selected. 
Typical exposure times (on source) are $\approx$8 min. 
A total of 6 and 5 giants have been 
observed in NGC~6569 and NGC~6624, respectively. Figures~\ref{6569data},~\ref{6624data} show the IR image and 
the colour\--magnitude diagram (CMD) \citep{val05,val04a} 
of the core region of NGC~6569 and NGC~6624, respectively. The position in the near\--IR CMD of the giant stars 
spectroscopically observed is also shown. Table~\ref{photparam} lists the absolute coordinates, intrinsic
J, K colours and bolometric magnitudes of the giant stars in our sample.

The two raw dimensional spectra were processed using the REDSPEC IDL\--based package written at the UCLA IR Laboratory. 
Each order has been sky subtracted by using the pairs of spectra taken with the object nodded along the slit, and subsequently 
flat\--field corrected. Wavelength calibration has been performed using arc lamp and a second\--order polynomial solution, while telluric 
features have been removed dividing by the featureless spectrum of an O star. 

The near\--IR spectra of cool stars are characterized by many CN, OH and CO molecular lines. 
At the NIRSPEC resolution $\rm R=25,000$ several single roto\--vibrational OH lines and CO bandheads can be measured to derive 
accurate oxygen and carbon abundances. Most of the CN molecular lines are, instead, faint and blended with the stronger CO, OH and atomic lines,
preventing any reliable abundance estimates of nitrogen.
Other metal abundances can be derived from the atomic lines of Fe~I, Mg~I, Si~I, Ti~I, Ca~I and 
Al~I.

Abundance analysis is performed combining full spectro\--synthesis techniques with equivalent width measurements of representative lines.
Using an updated version \citep{ori02} of the code described in \citet{ori93} we compute suitable synthetic spectra of giant stars.
The main features of the code and the overall spectral synthesis procedure have been widely discussed and tested 
in our previous papers and they will not be repeated here. 
Here we only stress that the code use the LTE approximation and is based on the molecular blanketed model atmospheres 
of \citet{jbk80} at temperatures $\leq 4000$~K and the ATLAS9 models for temperatures above 4000~K. 
The reference solar abundances are from \citet{gs98}.

A grid of model spectra is produced by varying the stellar parameters around the photometric values and the
abundances and abundance patterns over a large range.
Stellar temperatures are estimated from the (J\--K)$_0$ colours (see Tables~\ref{photparam}) and from molecular lines.
Gravity is derived from the theoretical evolutionary tracks according to the location of the stars on the red giant branch. An
average value $\xi$=2.0~km~s$^{-1}$ has been adopted for the microturbolence velocity \citep[see also][]{ori97}.
Tighter constrains on stellar parameters are obtained by the simultaneous spectral fitting of several CO and OH molecular bands, which
are very sensitive to variations of temperature, gravity and microturbolence.
As a figure of merit we adopt the difference between the model and the observed spectrum
(hereafter $\delta$). In order to quantify systematic discrepancies $\delta$ is more powerful than the classical
$\chi ^2$ test, which is instead equally sensitive to random and systematic scatters.
Equivalent widths of selected lines (see Table~\ref{ewpar}) are computed by Gaussian fitting the
line profiles with a $\leq$20\% overall uncertainty.
The model that best reproduces the overall observed spectrum is the same model that best reproduces the
equivalent widths of the selected lines and is chosen as the best-fit model.
Solutions with $\rm \Delta T_{eff} = \pm200$~K, $\rm \Delta log~g \pm0.5$~dex and $\rm \Delta \xi = \mp 0.5~km~s^{-1}$ and corresponding
$\pm$0.2~dex abundance variations from the best\--fitting one as well as solutions with $\pm$0.1~dex and  $\pm$0.2~dex abundance variations,
are typically significant at $1\leq \sigma \leq 3$ level only, using the figure of merit
mentioned above \citep[see][]{ori04}.
If the stellar parameters are varied simultaneously, appreciable variations in the spectrum can be measured for
smaller (by a factor of 2\--3) variations of the single quantities, depending on the range of temperature
and metallicity of the observed stars.
The adopted variation for the stellar parameters are somewhat conservative and have been empirically estimated
comparing observed spectra with synthetic ones covering a fine grid of stellar parameters.
At the observed spectral resolution (R=25,000) and signal\--to\--noise (30\--50), single variation of 100\--200~K in $T_{eff}$ or 0.5~dex in log~g or
0.5~km~s$^{-1}$ in $\xi$ produce variations in the spectrum which can be reliably measured/disentangled
from the observational errors.
Typically, for the observed red giant branch stars in metal\--rich globular clusters these variations can
introduce a maximum systematic uncertainty in the derived abundances between 0.1 and 0.2~dex. It must be noted, however, that
since the stellar features under consideration show a similar trend with variation in the
stellar parameters, although with different sensitivity, {\it relative} abundances are less dependent on the stellar parameter assumptions
(i.e. on the systematic errors) and their values are well constrained down to $\approx \pm$0.1~dex.

\begin{figure*} 
\includegraphics[width=8.5cm]{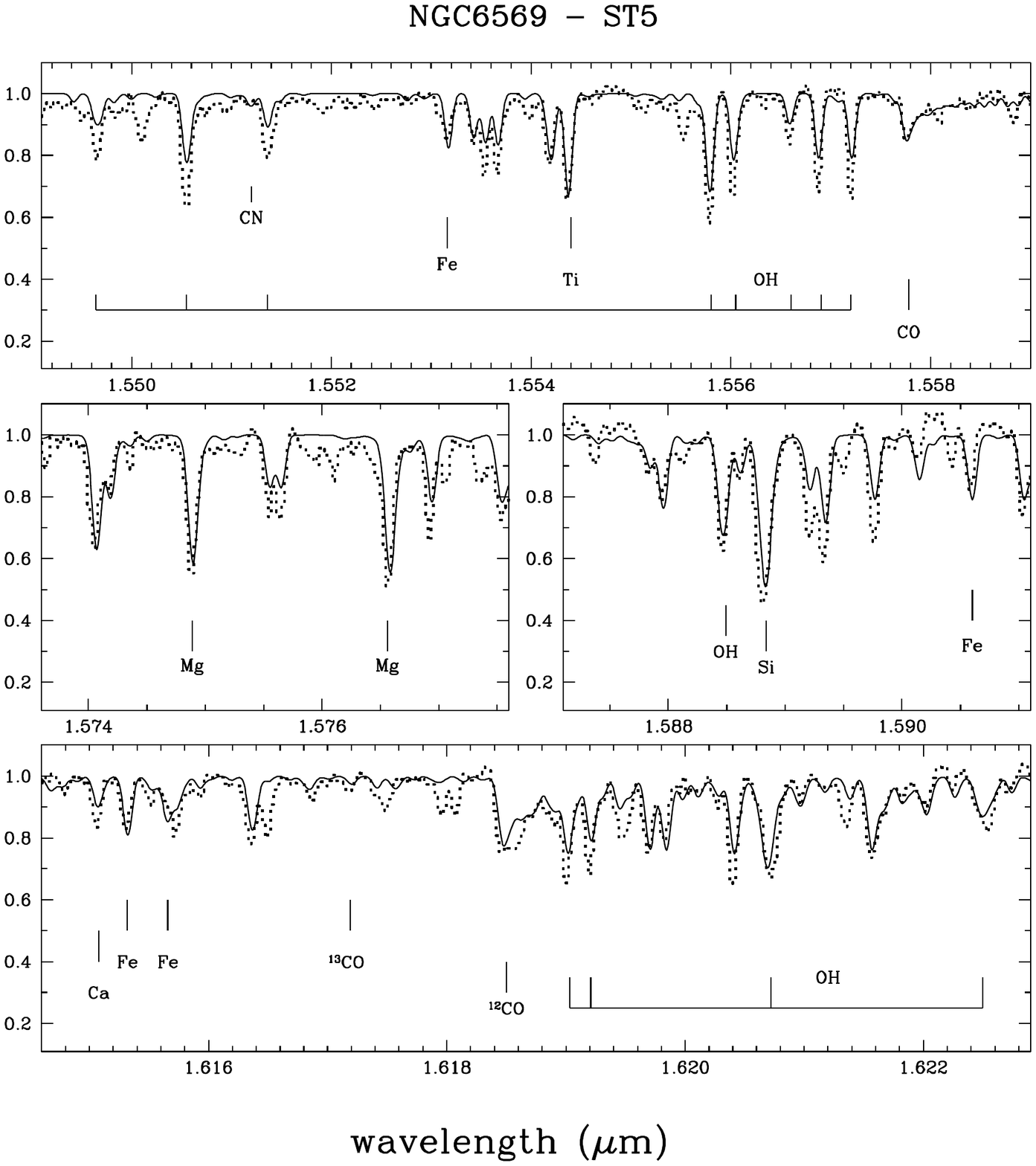}
\includegraphics[width=8.5cm]{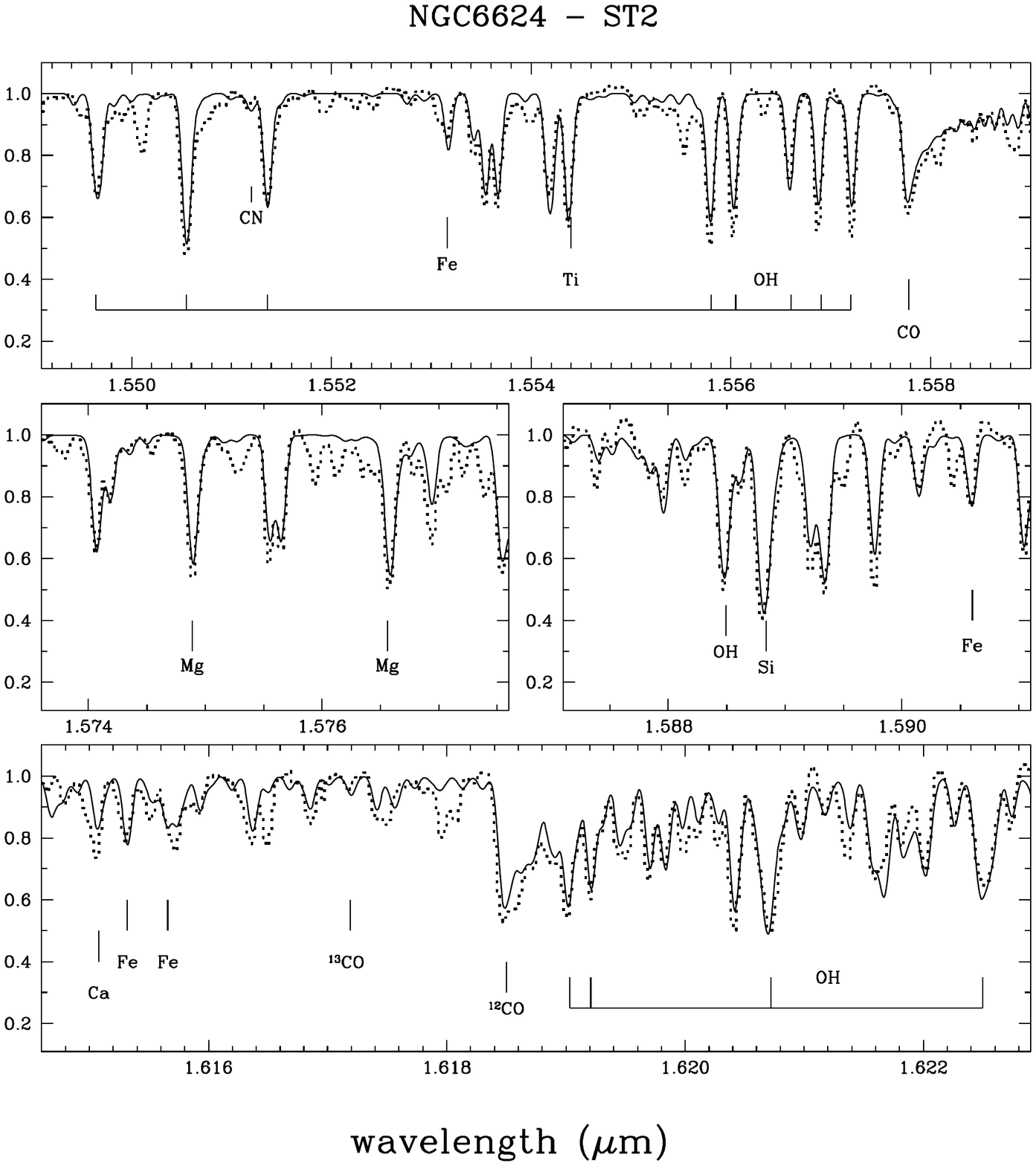}
\caption{Selected portions of the observed echelle spectra (dotted lines) of one giants in NGC~6569 ({\it left panel}) and in NGC~6624
({\it right panel}) with our best\--fitting synthetic spectrum (solid line) superimposed. A few important molecular and atomic lines of
slight interest are marked.}
\label{spec}
\end{figure*}

\begin{table*}
\begin{center}
\caption{Measured equivalent widths (in units of m\AA) of a few representative lines for the giants observed in NGC~6569 and NGC~6624}
\label{ewpar}
\begin{tabular}{lcccccccccccc}
\hline \hline
&&&&&&&&&&&&\\
&\multicolumn{6}{c}{NGC~6569}& &\multicolumn{5}{c}{NGC~6624}\\
&&&&&&&&&&&&\\
\hline \hline
Star & \#1 & \#2 & \#3 & \#4 & \#5 & \#6 & & \#1 & \#2 & \#3 & \#4 & \#5\\
&&&&&&&&&&&&\\
Ca~$\lambda $1.61508   & 161 &187  & 168 &  150 & 155 & 150 && 228 & 231 & 210  & 208 & 230\\
&&&&&&&&&&&&\\
Fe~$\lambda $1.61532   & 165 &170  & 171 &  154 & 151 & 168 && 184 & 183 & 192  & 198 & 190\\
&&&&&&&&&&&&\\
Fe~$\lambda $1.55317   & 138 &142  & 139 &  133 & 134 & 140 && 155 & 152 & 167  & 172 & 160\\
&&&&&&&&&&&&\\
Mg~$\lambda $1.57658   & 397 &395  & 387 &  395 & 404 & 396 && 405 & 393 & 415  & 410 & 406\\
&&&&&&&&&&&&\\
Si~$\lambda $1.58884   & 472 &510  & 473 &  477 & 468 & 457 && 497 & 508 & 495  & 476 & 515\\
&&&&&&&&&&&&\\
OH~$\lambda $1.55688   & 244 &326  & 236 &  240 & 245 & 253 && 304 & 339 & 263  & 245 & 340\\
&&&&&&&&&&&&\\
OH~$\lambda $1.55721   & 250 &330  & 243 &  243 & 248 & 256 && 307 & 348 & 270  & 256 & 356\\
&&&&&&&&&&&&\\
Ti~$\lambda $1.55437   & 281 &340  & 275 &  280 & 285 & 286 && 358 & 341 & 335  & 329 & 360\\
&&&&&&&&&&&&\\
Al~$\lambda $1.67633   & 291 &303  & 276 &  275 & 296 & 297 && 330 & 325 & 310  & 312 & 332\\
&&&&&&&&&&&&\\
\hline
\end{tabular}
\end{center}
\end{table*}

\begin{table*}
\begin{center}
\caption{Stellar parameters and abundances for the giants observed in NGC~6569 and NGC~6624}
\label{abund}
\begin{tabular}{lcccccccccccc}
\hline \hline
&&&&&&&&&&&&\\
&\multicolumn{6}{c}{NGC~6569}& &\multicolumn{5}{c}{NGC~6624}\\
&&&&&&&&&&&&\\
\hline \hline
Star & \#1 & \#2 & \#3 & \#4 & \#5 & \#6 & & \#1 & \#2 & \#3 & \#4 & \#5\\
&&&&&&&&&&&&\\
T$_{eff}$[K] &4000&3600&4000&4000&4000&4000&&3600&3600&3800&3800&3600\\
&&&&&&&&&&&&\\
log~g &1.0&0.5&1.0&1.0&1.0&1.0&&0.5&0.5&0.5&0.5&0.5\\
&&&&&&&&&&&&\\
v$^a_r $[km~s$^{-1]}$ &-33&-49&-42&-48&-56&-54&&51&48&45&53&60\\
&&&&&&&&&&&&\\
${\rm [Fe/H]}$ &-0.75&-0.81&-0.75&-0.85&-0.84&-0.76&&-0.72&-0.73&-0.65&-0.64&-0.69\\
 & $\pm$0.10 & $\pm$0.07 & $\pm$0.09 & $\pm$0.09 & $\pm$0.09 & $\pm$0.10 & & $\pm$0.08 & $\pm$0.09& $\pm$0.10 & $\pm$0.10 & $\pm$0.09\\
&&&&&&&&&&&&\\
${\rm [O/Fe]}$ &+0.50&+0.37&+0.37&+0.58&+0.52&+0.53&&+0.38&+0.60&+0.37&+0.22&+0.49\\
 & $\pm$0.10 & $\pm$0.09 & $\pm$0.10 & $\pm$0.10 & $\pm$0.09 & $\pm$0.10 & & $\pm$0.10& $\pm$0.12& $\pm$0.11 & $\pm$0.10 & $\pm$0.11 \\
&&&&&&&&&&&&\\
${\rm [Ca/Fe]}$ &+0.35&+0.28&+0.35&+0.35&+0.29&+0.25&&+0.42&+0.43&+0.35&+0.35&+0.43\\
 & $\pm$0.16 & $\pm$0.13 & $\pm$0.15 & $\pm$0.16 & $\pm$0.15 & $\pm$0.16 & & $\pm$0.14& $\pm$0.14 & $\pm$0.15& $\pm$0.15 & $\pm$0.14 \\
&&&&&&&&&&&&\\
${\rm [Si/Fe]}$ &+0.47&+0.53&+0.48&+0.56&+0.56&+0.36&&+0.42&+0.43&+0.35&+0.24&+0.48\\
 & $\pm$0.16 & $\pm$0.19& $\pm$0.16& $\pm$0.16 & $\pm$0.16& $\pm$0.16 & & $\pm$0.19& $\pm$0.25 & $\pm$0.20 & $\pm$0.19& $\pm$0.20 \\
&&&&&&&&&&&&\\
${\rm [Mg/Fe]}$ &+0.49&+0.54&+0.40&+0.55&+0.56&+0.46&&+0.50&+0.43&+0.42&+0.37&+0.39\\
 & $\pm$ 0.17& $\pm$0.16& $\pm$0.17& $\pm$0.16 & $\pm$0.16& $\pm$0.17 & & $\pm$0.15& $\pm$0.16 & $\pm$0.17 & $\pm$0.19 & $\pm$0.15 \\
&&&&&&&&&&&&\\
${\rm [Ti/Fe]}$ &+0.35&+0.41&+0.35&+0.45&+0.44&+0.41&&+0.42&+0.33&+0.35&+0.34&+0.39\\
 & $\pm$ 0.19& $\pm$0.16 & $\pm$0.19& $\pm$0.19 & $\pm$0.18 & $\pm$0.19& & $\pm$0.15& $\pm$0.16 & $\pm$0.18 & $\pm$0.20 & $\pm$0.16 \\
&&&&&&&&&&&&\\
${\rm [Al/Fe]}$ &+0.35&+0.31&+0.25&+0.35&+0.54&+0.46&&+0.42&+0.38&+0.40&+0.38&+0.39\\
 & $\pm$ 0.20& $\pm$0.17 & $\pm$0.18& $\pm$0.20 & $\pm$0.18 & $\pm$0.20& & $\pm$0.14& $\pm$0.14 & $\pm$0.15 & $\pm$0.15 & $\pm$0.14 \\
&&&&&&&&&&&&\\
${\rm [C/Fe]} $&-0.45&-0.19&-0.25&-0.35&-0.16&-0.24&&-0.18&-0.17&-0.45&-0.56&-0.11\\
 & $\pm$ 0.12& $\pm$0.10 & $\pm$0.12 & $\pm$0.12 & $\pm$0.11 & $\pm$0.12 & & $\pm$0.11& $\pm$0.12 & $\pm$0.12 & $\pm$0.12 & $\pm$0.11 \\
\hline
\multicolumn{13}{l} {(a) Heliocentric radial velocity.}\\
\multicolumn{13}{l} {The random error due to EW measurement is quoted for each element abundance.}\\
\\
\end{tabular}
\end{center}
\end{table*}


\section {Results}
\subsection {NGC~6569}
Photometric stellar temperature estimates and bolometric magnitudes have been computed by using 
the near\--IR photometry by \citet{val05}, their cluster reddening (E(B \-- V)=0.49) and distance ((m \-- M)$_0$=15.40) values, and 
the colour\--temperature
transformations and bolometric correction of \citet{mont98}, specifically calibrated for globular cluster giants.
We constrain bolometric magnitudes in the range-3.0$\rm \leq M_{bol}\leq$-2.8 (see Table~\ref{photparam}). 
The final adopted temperatures, listed in Table~\ref{abund}, are then derived by fitting the CO and, in particular, the OH bands,
which are especially temperature sensitive in cool giants.

Figure~\ref{spec}, left panel, shows our synthetic best\--fitting models superimposed on the observed spectrum of one out of six
observed giants in NGC~6569. For this cluster we find an average heliocentric radial velocity of $\rm <v_r>=-47\pm 4~km~s^{-1}$ and 
a velocity dispersion of $\rm \approx 8~km~s^{-1}$. Our radial velocity estimate does not agree with the available literature value
reported by \citet{harris} (i.e. -28.1~$\rm km~s^{-1}$) based on optical integrated spectroscopic measurement performed by \citet{6569rv}.
A possible reason for such discrepancy could be the presence of bright foreground field stars along the cluster core line of sight,
which could eventually dominate the result of the integrated spectrum. 
Indeed, the [V, V\--I] CMD of the innermost cluster region ($\rm r<1\arcmin$, see Fig.~4 of \citet{ort01}) 
shows the presence of 3 field stars
brighter than the RGB tip. 
From our overall spectra analysis, we find average [Fe/H]$=-0.79\pm 0.02$~dex, [O/Fe]$=+0.48\pm 0.04$~dex, 
[$\alpha$/Fe]$=+0.43\pm 0.02$~dex, [Al/Fe]$=+0.38\pm 0.10$~dex, and a carbon depletion of [C/Fe]$=-0.27\pm0.04$~dex.
Low $\rm ^{12}C/^{13}C\le 7$ have been also measured.

\subsection {NGC~6624}
The near\--IR photometry of \citet{val04a} and their derived $E(B-V)=0.28$ reddening and $\rm (m-M)_0=14.63$ distance modulus have 
been used to derive the photometric estimates of 
the stellar temperatures and bolometric magnitudes for the giants observed in NGC~6624. 
As shown in Table~\ref{photparam}, these stars have bolometric magnitude in the range
$\rm -3.0\leq M_{bol} \leq -3.8$.

The giants in our sample are likely to be cluster members with an average heliocentric radial velocity of $\rm <v_r>=+51\pm 3~km~s^{-1}$
and a velocity dispersion $\rm \sigma \approx 6~km~s^{-1}$, in agreement with the value ($\rm <v_r>=+54.3~km~s^{-1}$) 
listed by \citet{harris}.

For this cluster our abundances analysis (see Table~\ref{abund}) gives an average [Fe/H]$=-0.69\pm 0.02$~dex, 
[O/Fe]$=+0.41\pm 0.06$~dex, [$\alpha$/Fe]$=+0.39\pm 0.02$~dex, [Al/Fe]$=+0.39\pm 0.02$~dex,
and a carbon depletion of [C/Fe]$=-0.29\pm0.08$~dex.
Low $\rm ^{12}C/^{13}C\approx 5$ have been also measured.

\section {Discussion and Conclusions}

The [$\alpha$/Fe] enhancement (where $\alpha = < Ti, Si, Mg, Ca>$) measured in both NGC~6569 and NGC~6624 is consistent 
with those measured in the other 10 GCs
and in the four inner Bulge fields surveyed by our group.
Moreover, an overall agreement is also found with the results obtained so far in outer regions along the minor axis 
from \--3$^{\circ}$ to \--12$^{\circ}$ galactic latitude by \citet{lec07,alvbrit10,plaut,oscar10}.
All this measurements indicate that the bulk of the Bulge population as a whole, likely formed from a gas  
mainly enriched by type II SNe on a short timescale, before substantial explosion of type Ia SNe and at early epochs, 
as suggested by old age \citep[$\geq 10$Gy][]{kuirmr02,zocc03,sahu06}

Also the [C/Fe] depletion and the low $\rm ^{12}C/^{13}C$ are consistent with previous measurements of Bulge giants and 
confirming the presence of extra\--mixing process during the evolution along the RGB in metal\--rich environments.

The relatively small number of giants observed in NGC~6569 and NGC~6624 do not allow us to properly check 
for possible Al-O and Al-Mg anti-correlations.

\section*{Acknowledgments}

RMR acknowledges support from grant number AST\--0709479 from the National Science Foundation. 
LO acknowledges financial support by the
Ministero dell'Istru\-zio\-ne, Universit\`a e Ricerca (MIUR) and the
Istitututo Nazionale di Astrofisica.
The authors are grateful to the staff at the Keck Observatory and to Ian McLean and the NIRSPEC team. 
The authors wish to recognize and acknowledge the very significant cultural role and reverence that the 
summit of Mauna Kea has always had within the indigenous Hawaiian community. We are most fortunate to 
have the opportunity to conduct observations from this mountain.


\bsp

\label{lastpage}

\end{document}